# Pressure-induced anomalous enhancement in superconducting critical temperature of transition-metal chalcogenide $Ta_2PdS_6$ and $Ta_2PdSe_6$

*Ryo Matsumoto[1], Akitoshi Nakano[3], Takafumi D Yamamoto[4], Kensei Terashima[1], Kazuki Yamane[1,2], Masahiro Ohkuma[1], Ichiro Terasaki[3], Yoshihiko Takano[1,2]

[1]Research Center for Materials Nanoarchitectonics (MANA),
National Institute for Materials Science, Tsukuba, Ibaraki 305-0047, Japan
[2]Graduate School of Pure and Applied Sciences, University of Tsukuba, Tsukuba, Ibaraki 305-8577, Japan
[3]Department of Physics, Nagoya University, Nagoya 464-8602, Japan
[4] Department of Material Science and Technology, Tokyo University of Science, Tokyo 125-8585, Japan

*Corresponding author; Email: MATSUMOTO.Ryo@nims.go.jp

**Abstract**

The emergence of a second dome in the superconducting phase through pressure-driven manipulation of crystal structures in materials has attracted considerable attention. Transition metal chalcogenides (TMCs) represent a highly promising platform, as the second dome has been observed in several binary compounds. Recently, ternary TMCs such as $Ta_2PdS_6$ and $Ta_2PdSe_6$ have exhibited pressure-induced superconducting domes. In this study, we perform electrical transport measurements of $Ta_2PdS_6$ and $Ta_2PdSe_6$ under extremely high pressures exceeding 100 GPa, namely uninvestigated regions in previous reports, to reveal the emergence of the second dome. The superconducting critical temperatures ($T_c$) in both $Ta_2PdS_6$ and $Ta_2PdSe_6$ initially decrease with increasing pressure. Subsequently, the $T_c$s tend to enhance drastically above 100 GPa. Notably, the maximum $T_c$ in $Ta_2PdS_6$ is 11.2 K at 130.0 GPa, which is a relatively high record among the TMCs. The emergence of the second dome in $Ta_2PdS_6$ and $Ta_2PdSe_6$ opens further motivation for the investigation under extreme conditions beyond the first dome to find hidden ordered phases.

# 1. Introduction

Application of high pressure is an effective method to tune the structural and electronic properties of materials. In particular, the effects of pressure on the emergence and suppression of ordered phases, such as charge density wave (CDW), spin density wave (SDW), and superconductivity (SC), have attracted significant research attention in recent decades [1–6]. The transition temperatures of these ordered phases typically exhibit a dome-like behavior under applied pressure. Interestingly, the emergence of a second SC phase is observed in several materials following the suppression of the first SC dome [7]. Although most second SC domes show a lower or comparable $T_c$ than that of the first SC dome, certain compounds, such as the ion-based superconductors, exhibit higher $T_c$ in the second dome [8,9]. The reemergent superconductivity with a higher $T_c$ in the second dome of other material families has become a focused research area, as theoretical calculations have predicted in high-$T_c$ materials [10].

Transition metal chalcogenides (TMCs) have been focused as the platform for studying dome-like behavior in ordered phases under high pressure, particularly in binary systems, such as a pressure-driven suppression of the CDW phase and the emergence of SC phase [11–13]. Recently, ternary TMCs of $Ta_2PdS_6$ and $Ta_2PdSe_6$, which crystallize in a monoclinic quasi-1D structure with space group $C2/m$, as shown in Fig. S1, have been studied actively due to their anomalous electrical properties at ambient pressure. For instance, $Ta_2PdS_6$ exhibits semiconducting behavior with an electron carrier density of $4.6 \times 10^{18}$ cm$^{-3}$ at 100 K [14], despite band calculations predicting a metallic state in the electronic structure. The transport measurements of the semimetal $Ta_2PdSe_6$ reveal a non-Fermi liquid-like temperature dependence, indicating an exotic electronic state with an unconventional scattering process for charge carriers [15]. Among the investigations of fundamental physics in this system, the emergence of the SC phase through the application of pressure has been reported in both $Ta_2PdS_6$ and $Ta_2PdSe_6$, with dome-like behavior of $T_c$ [16,17]. However, the question of reemergent SC phases beyond the first dome in this system remains an open issue.

In this study, we perform electrical transport measurements on single-crystalline $Ta_2PdS_6$ and $Ta_2PdSe_6$ under extremely high pressures to investigate the SC properties beyond the first dome. Through the electrical resistance measurements on $Ta_2PdS_6$ and $Ta_2PdSe_6$ above the megabar pressure range, we reveal the existence of a second SC phase with a $T_c$ higher than that in the first dome. Specifically, the $T_c$ in $Ta_2PdS_6$ rises drastically above 80.4 GPa, reaching 11.2 K at 130.0 GPa. This observation of the second SC dome with a higher $T_c$ under extreme pressure offers significant motivation to investigate the megabar pressure region in functional materials, potentially opening new aspects of materials physics.

# 2. Materials and methods

High-quality single crystals of $Ta_2PdS_6$ and $Ta_2PdSe_6$ were grown for electrical transport measurements and Raman spectroscopy under high pressure using a chemical vapor transport with a transport agent of $I_2$ by referring to previous reports [14,18,19]. Starting materials of tantalum (99.9%), palladium (99.9%), and sulfur (99.999%) or selenium (99.9% or 99.999%) were loaded into an evacuated quartz tube with an $I_2$ concentration of ~3 mg cm$^{-3}$. A temperature difference of 145°C between 875 and 730°C in a three-zone furnace facilitated crystal growth over four days. The compositional ratio is evaluated by energy dispersive spectrometry (EDX) using a JSM-6010LA (JEOL). Details of the characterization of the obtained sample at ambient pressure were provided elsewhere [14]. A polycrystalline sample of $Ta_2PdS_6$ was synthesized via a solid-state reaction for a structural analysis under high pressure. The same starting powders as the single crystalline sample were used. The raw powder, once heated to 550°C, was re-grounded and heated at 730°C for 48 h in a tube furnace. The obtained sample was identified as a single phase of $Ta_2PdS_6$ by powder X-ray diffraction (XRD). The crystal structure was displayed by VESTA software [20].

Electrical transport measurements under high pressure in $Ta_2PdS_6$ and $Ta_2PdSe_6$ were conducted within a diamond anvil cell (DAC), employing a diamond electrode [21–23]. The temperature ($T$) dependence of resistance ($R$) was measured in the physical property measurement system (PPMS, Quantum Design) with a superconducting magnet. Raman spectroscopy was also performed at room temperature for the sample to evaluate the vibrational modes under high pressure. The diamond anvil equipped a beveled culet with a diameter of ~100 μm. A cleaved single-crystalline sample was positioned onto the diamond anvil with the electrodes, and a SUS316 plate served as a metal gasket. The pressure-transmitting medium and insulating layer consisted of cubic BN powders. The applied pressure was estimated using fluorescence from ruby powder placed on the culet [24] and the Raman spectrum from the diamond anvil itself [25], employing an inVia Raman Microscope (RENISHAW).

The crystal structure of $Ta_2PdS_6$ under high pressure was investigated through XRD measurements in the DAC. The culet of diamond anvil was 300 μm, the gasket was a tungsten plate, and the PTM was the sample itself. These measurements were carried out using synchrotron radiation at the AR-NE1A beamline of the Photon Factory (PF) located at the High Energy Accelerator Research Organization (KEK). The X-ray beam monochromatized to an energy of 30 keV ($\lambda = 0.4175$ Å), was directed to the sample in the DAC through a collimator with a diameter of 50 μm. The obtained XRD patterns were integrated into a one-dimensional profile using IPAnalyzer [26]. The applied pressure was estimated using the same procedures as those used for electrical measurements.

## 3. Results and discussion

Figure 1 shows the *R-T* characteristics of $Ta_2PdS_6$ under various pressures up to (a) 25.6 GPa, (b) 130.0 GPa, and (c) enlarged plots at low temperatures. The behavior of $Ta_2PdS_6$ under pressure is divided into three regions: (i) suppression of semiconducting behavior and transition to metallic property below 21.5 GPa, (ii) emergence of SC phase at 25.6 GPa with a gradual change in $T_c$ up to 70.0 GPa, and (iii) a drastic enhancement in $T_c$ above 80.4 GPa. In region (i), $Ta_2PdS_6$ exhibits semiconducting characteristics, with a negative $dR/dT$ at low temperatures under the lowest pressure of 1.4 GPa. This semiconducting curve is gradually suppressed with increasing pressure up to 18.6 GPa. Conversely, the $R$-$T$ properties show metallic behavior across the measured temperature range at 21.5 GPa, indicating a semiconductor-to-metal transition. According to in-situ XRD analysis and Raman spectroscopy under pressure (Fig. S2), this metallization is due to a pressure-induced isosymmetric structural transition in $Ta_2PdS_6$ [17]. In region (ii), a sharp decrease in $R$, corresponding to the emergence of SC phase, appears above 25.6 GPa. The $R$ starts to drop from 6.5 K, defined as $T_c^{onset}$, and gradually decreases to zero with several kinks, indicating inhomogeneous superconductivity. At 58.9 GPa, $R$ reaches zero at $T_c^{zero}$ of 2.5 K, with only slight changes in $T_c$ as pressure increases up to 70.0 GPa. In this SC region, a sign reversal in Hall resistivity is reported in both $Ta_2PdSe_6$ [16] and $Ta_2PdS_6$ [17]. However, in our observation for $Ta_2PdS_6$, a negative slope in Hall resistance, due to electron carriers, is dominant even at the lowest pressure and maintained up to SC region, as indicated in Fig. S3. Notably, $T_c$ increases sharply under further compression above 80.4 GPa, corresponding to region (iii). Although the rate of increase in $T_c$ slows above 100.3 GPa, the enhancement does not fully saturate even at the highest pressure, reaching maximum values of 11.2 K for $T_c^{onset}$ and 8.5 K for $T_c^{zero}$ at 130.0 GPa, marking a relatively high $T_c$ among TMCs.

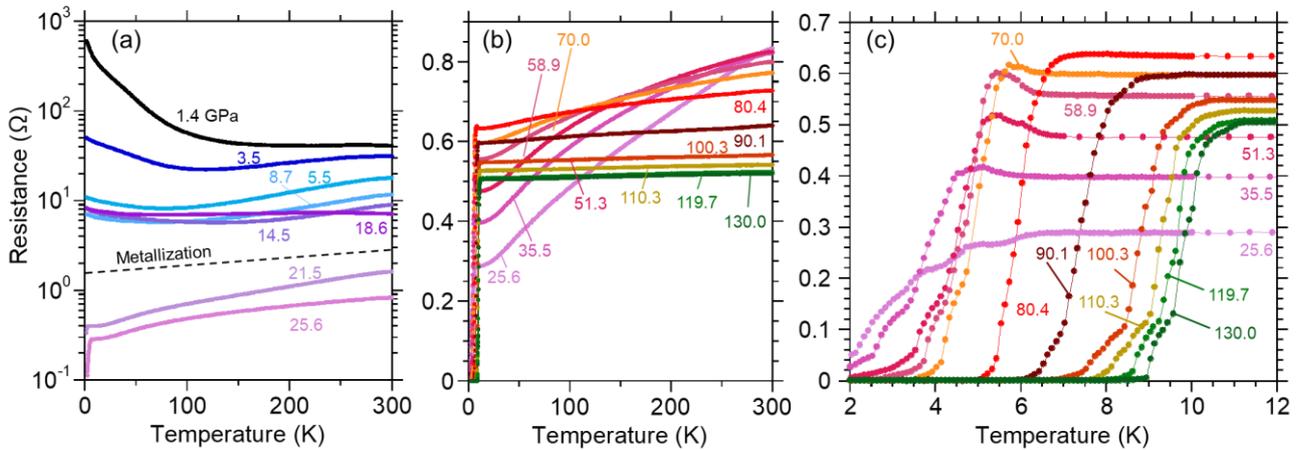

**Fig. 1. (a) Temperature dependence of resistance under various pressures from 1.4 to 25.6 GPa and (b) 25.6 to 130.0 GPa in $Ta_2PdS_6$. (c) Enlarged plots around superconducting transition.**

Figure 2 presents the $R$-$T$ properties of $Ta_2PdSe_6$ under pressures up to (a) 60.5 GPa, (b) 118.8 GPa, and (c) enlarged plots at low temperatures. The behavior of $Ta_2PdSe_6$ under pressure is also divided into two regions: (i) emergence of pressure-induced SC at 16.9 GPa and gradual change in $T_c$ up to 92.1 GPa, and (ii) drastic enhancement in $T_c$ above 102.1 GPa. In contrast to $Ta_2PdS_6$, $Ta_2PdSe_6$ exhibits metallic behavior even at low pressures, and its SC phase appears at a lower pressure of 16.9 GPa. The SC transition is initially broad at 16.9 GPa due to pressure inhomogeneity. With applying pressure, $T_c^{onset}$ decreases gradually, and the SC transition becomes sharp, with zero resistance achieved above 60.5 GPa. With further compression, $T_c^{onset}$ shows a slight increase up to 102.1 GPa. A drastic rise in $T_c^{onset}$ above 102.1 GPa indicates the emergence of a second dome of pressure-induced SC in $Ta_2PdSe_6$.

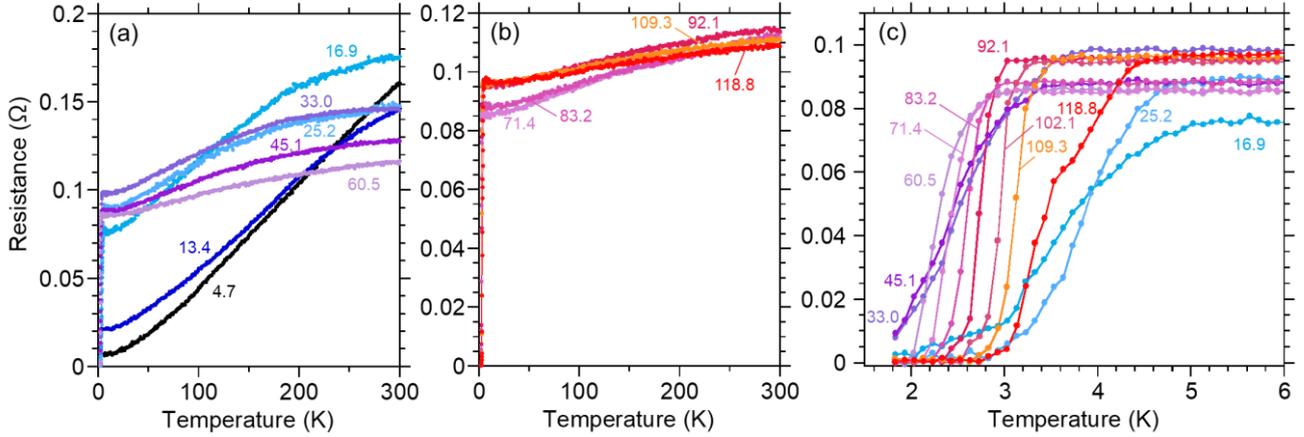

**Fig. 2. (a) Temperature dependence of resistance under various pressures from 4.7 to 60.5 GPa and (b) 71.4 to 118.8 GPa in $Ta_2PdSe_6$. (c) Enlarged plots around superconducting transition.**

Figure 3 presents the pressure-dependent $T_c^{onset}$ in $Ta_2PdS_6$ and $Ta_2PdSe_6$ up to the megabar region, with comparisons to previous reports. In the unexplored pressure range beyond previous studies on $Ta_2PdS_6$ and $Ta_2PdSe_6$, specifically above the megabar region, our samples show a significant enhancement of $T_c$, revealing the existence of the second SC dome. Even at the maximum pressures in this study, the increases in $T_c$ are not fully saturated. In contrast to most two-dome SC systems, where the second dome exhibits a lower or comparable $T_c$ (e.g., $CsV_3Sb_5$ [7]), $Ta_2PdS_6$ shows double $T_c$ of 11.2 K at 130.0 GPa than that observed at 70.0 GPa. Similarly, $T_c$ in $Ta_2PdSe_6$ at 118.8 GPa is comparable to that at 16.9 GPa and continues to enhance steeply with applying pressure. The observed higher $T_c$s in the second SC domes are unique features in these materials. Although the behavior of $Ta_2PdS_6$ in low-pressure region differs from the previous report [17], a possible reason is slight changes in the composition. In the previous report, the composition was Ta:Pd:S = 1.9 : 1 : 5.7 with a p-type carrier [17], whereas our sample is Ta:Pd:S = 2.0 : 1 : 5.7 with an n-type carrier as shown by Hall measurement (Fig. S3). The slight difference may induce the distinct behavior at lower pressures, as metal vacancies introduce p-type doping in several TMCs [32,33].

This drastic enhancement of $T_c$ under high pressure is unusual, as the application of pressure typically decreases $T_c$ due to phonon hardening and a reduction in the electronic density of state (DOS) at Fermi energy, based on Bardeen–Cooper–Schrieffer (BCS) theory [27,28]. Two-dome behavior of $T_c$ is typically associated with structural phase transitions, as seen in FeS [29], $LaFeAsO_{1-x}F_x$ [30], and other compounds [31]. However, a recently discovered SC phase in the topological kagome metal $CsV_3Sb_5$ shows two-dome $T_c$ attributed to a pressure-induced Lifshitz transition, resulting from a reconstruction of the Fermi surface without structural phase transition [7]. The elevated $T_c$ in $Ta_2PdS_6$ and $Ta_2PdSe_6$ likely relate to a modification of electronic structure rather than a structural phase transition, as discussed later.

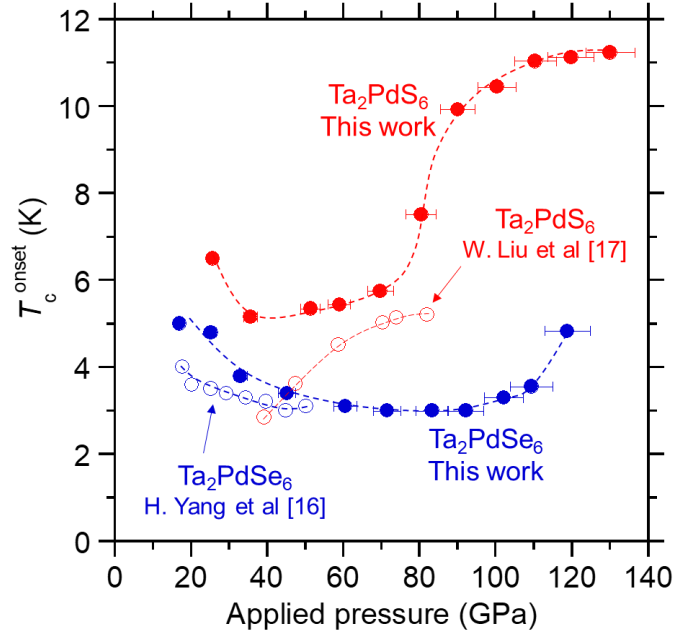

**Fig. 3 Pressure-dependent $T_c$ in $Ta_2PdS_6$ and $Ta_2PdSe_6$ up to megabar region with a comparison of previously reported data of $Ta_2PdS_6$ [17] and $Ta_2PdSe_6$ [16]. The dashed lines are the guide for the eyes.**

To discuss the origin of anomalous enhancements, the pressure dependence of $R$ at 300 K ($R_{300K}$) in both compounds is plotted in Fig. 4 (a). In $Ta_2PdS_6$, $R_{300K}$ decreases monotonically with increasing pressure and reduces discretely at 21.5 GPa near $P_c$, suggesting a first-order transition. $R_{300K}$ gradually decreases further above 25.6 GPa until the maximum pressure of 130.0 GPa, without any additional discrete shifts. Similar gradual changes in $R_{300K}$ are observed in $Ta_2PdSe_6$. Figure 4 (b) plots the pressure dependence of the upper critical field $\mu_0H_{c2}(0)$ and coherence length at zero temperature $\xi(0)$ in $Ta_2PdS_6$ and $Ta_2PdSe_6$. To estimate these parameters, $R$-$T$ curves under various magnetic fields were fitted with the Werthamer-Helfand-Hohenberg (WHH) model [34,35] using a $T_c$ criterion at 95% of the normal resistance (Fig. S4). The maximum $\mu_0H_{c2}(0)$ is 11.3 T at 130.0 GPa in $Ta_2PdS_6$ and 3.5 T at 109.35 GPa in $Ta_2PdSe_6$. These values are lower than the weak-coupling Pauli limit ($1.84T_c$), suggesting the absence of the Pauli paramagnetic pair-breaking effect. The $\xi(0)$ was derived from the Ginzburg-Landau (GL) formula $\mu_0H_{c2}(0) = \Phi_0/2\pi\xi(0)^2$, where $\Phi_0$ is the fluxoid. The smooth changes in $R_{300K}$, $\mu_0H_{c2}(0)$, and $\xi(0)$ as pressure increases beyond $P_c$ suggest that the origin of the second SC dome is linked to modifications in electronic structures, such as a Lifshitz transition, as seen in kagome metal [7]. First-principles calculations on $Ta_2PdS_6$ indicate that enhanced DOS at Fermi energy plays a crucial role in the emergence of SC phase [17]. The observed $T_c$ enhancement is possibly associated with DOS peaks near Fermi energy, which is a favorable condition for superconductivity, as seen in high-$T_c$ hydrides [36]. Our finding of the hidden SC phase in $Ta_2PdS_6$ and $Ta_2PdSe_6$ reveals the importance of exploring extreme conditions to accelerate further development of materials physics.

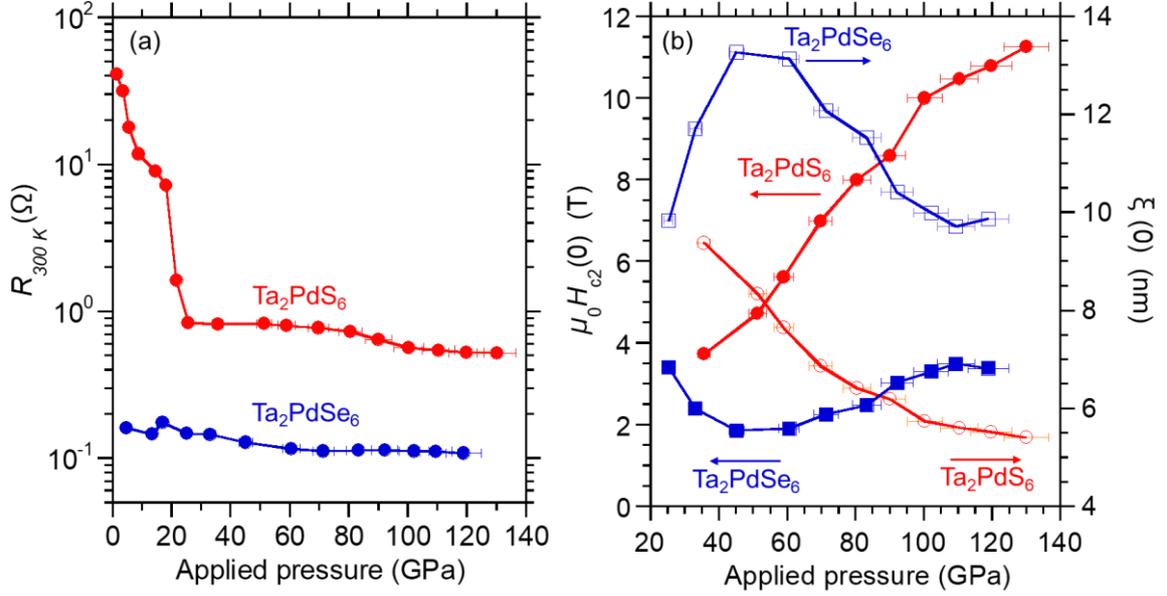

**Fig. 4 (a) Pressure-dependent resistance at 300 K, (b) $\mu_0H_{c2}(0)$, and $\xi(0)$ in $Ta_2PdS_6$ and $Ta_2PdSe_6$.**

## 4. Conclusion

In this study, we conduct the electrical transport measurements under high pressure up to 130.0 GPa in $Ta_2PdS_6$ and 118.8 GPa in $Ta_2PdSe_6$ to investigate SC properties beyond the first dome. The $T_c$ in $Ta_2PdS_6$ exhibits a drastic enhancement above 80.4 GPa and reaches 11.2 K at 130.0 GPa, which is almost twice the $T_c$ observed in the first dome. In $Ta_2PdSe_6$, the emergence of the second dome is also observed above 102.1 GPa. The observed higher $T_c$ in the second dome is motivative for further investigation of fundamental physics in this materials system. Although we consider that the drastic enhancements of $T_c$ in these compounds likely originate from modifications in the electronic structures, challenges remain in conducting structural analysis at the corresponding pressure.

**ACKNOWLEDGMENTS**

This work was partly supported by JSPS KAKENHI Grant Numbers 23H01835, 23K13549, and 23KK0088. The fabrication process of diamond electrodes was partially supported by the NIMS Nanofabrication Platform in the Nanotechnology Platform Project sponsored by the Ministry of Education, Culture, Sports, Science and Technology (MEXT), Japan. The synchrotron X-ray experiments were performed at AR-NE1A (KEK-PF) under the approval of Proposal No. 2022G049 with support from Dr. Y. Shibazaki (KEK).